\input{aipcheck}

\documentclass[
    ,final            % use final for the camera ready runs
  ]
  {aipproc}

\layoutstyle{6x9}

\begin{document}

\title{Structure Functions - Status and Prospect}

\author{J. C. Peng}{
  address={University of Illinois, Urbana, Illinois, 61801, U.S.A.}
}

\begin{abstract}
Current status and future prospects of the structure functions and
parton distribution studies are presented.
\end{abstract}

\maketitle

%%%%%%%%%%%%%%%%%%%%%%%%%%%%%%%%%%%%%%%%%%%%
%% MAINMATTER
%%%%%%%%%%%%%%%%%%%%%%%%%%%%%%%%%%%%%%%%%%%%

\section{Introduction}

The study of nucleon's structure functions and parton distributions
is an active area of research in nuclear and particle physics. 
The parton distributions address both the perturbative and 
nonperturbative aspects of QCD, and they also provide an essential 
input for describing hard processes in high-energy hadron
collisions. As a result of several decades's intense effort, the
unpolarized proton structure functions have been well mapped out over
a broad range of $Q^2$ and Bjorken-$x$. While these
data are invaluable for testing QCD and for extracting various parton
distributions, several questions remain unanswered. For example, the
unexpected finding of the flavor asymmetry of the light-quark sea ($\bar u,
\bar d$) suggests that other aspects of the flavor structure, such as
possible asymmetry between the $s$ and $\bar s$ sea quark distributions
and the bahavior of valence $d/u$ ratio at large $x$, need to be examined.
The issue of quark-hadron duality, as reflected in the intriguing 
similarity between the structure functions measured at the resonance
region and at the DIS region, also 
requires further studies.

Remarkable progress in the study of spin-dependent structure functions has 
been made since the discovery of the ``proton spin puzzle" in the 
late 1980's. Very active spin-physics programs have been pursued 
at many facilities including SLAC, CERN, HERA, JLab, and RHIC. 
The polarized DIS data now cover a
sufficiently broad $Q^2$ range for scaling-violation to be observed.
In recent years, new experimental tools such as semi-inclusive polarized 
DIS, polarized proton-proton collision, and deeply exclusive reactions
have been employed to address the major unresolved question in spin 
physics: How is the
proton's spin distributed among its various constituents? 

On the theory front, the formulation of the generalized parton distributions
as well as the identification of various $k_T$ (intrinsic transverse
momentum of partons)-dependent structure and fragmentation
functions have opened exciting new directions of research. Furthermore,
important progress in the Lattice calculations for the moments of various
parton distributions and in the extrapolations to their chiral limits
has been made.

In this review I will focus on recent progress in the following
areas:

\begin{itemize}

\item Flavor structure of parton distributions

\item Transition from high-$Q^2$ to low-$Q^2$

\item Novel distribution and fragmentation functions; Generalized parton distributions

\end{itemize}

\section{Flavor structures of parton distributions}
\subsection{$\bar d / \bar u$ flavor asymmetry}
The earliest parton models assumed that the proton sea was flavor symmetric,
even though the valence quark distributions are clearly flavor asymmetric.
The flavor symmetry assumption was not based on any known physics, and
it remained to be tested.
Under the assumption of a $\bar u$, $\bar d$ flavor-symmetric sea in
the nucleon, the Gottfried Sum Rule~\cite{gott}, $I_G =
\int_0^1 (F^p_2 (x,Q^2) - F^n_2 (x,Q^2))/x~ dx = 1/3$, 
is obtained. The NMC collaboration determined the Gottfried integral to be
$ 0.235\pm 0.026$, significantly below 1/3. This surprising result
can be explained by a large flavor asymmetry between the $\bar u$ 
and the $\bar d$.

The $x$ dependence of $\bar d / \bar u$ asymmetry has been determined by
proton-induced Drell-Yan (DY) as well as semi-inclusive DIS measurements.
Figure 1 shows that the Fermilab E866~\cite{towell01} 
DY cross section per nucleon for
$p + d$ clearly exceeds $p + p$, and it indicates an excess of $\bar d$
with respect to $\bar u$ over an appreciable range in $x$.
In contrast, the $\sigma(p+d)/2\sigma(p+p)$ ratios for $J/\Psi$ and
$\Upsilon$ production, also shown in Fig. 1, are very close to unity.
This reflects the dominance of gluon-gluon fusion process for
quarkonium production and the expectation that the gluon distributions
in the proton and in the neutron are identical.

\begin{figure}
\resizebox{1.0\textwidth}{!}{%
  \includegraphics{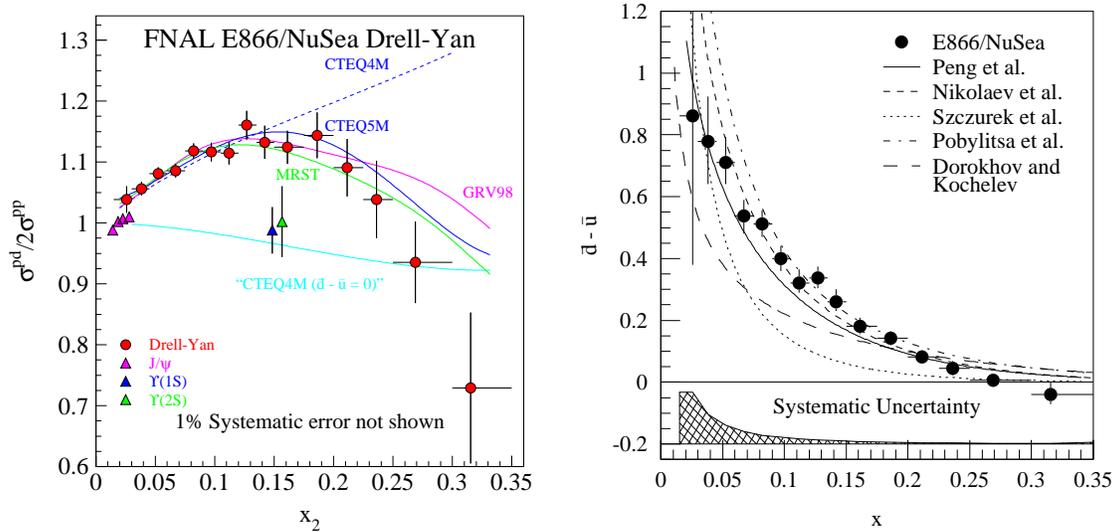}
}
\vspace*{-0.8in}
\caption{Left panel: Cross section ratios 
of $p+d$ over $2(p+p)$ for Drell-Yan, $J/\Psi$,
and $\Upsilon$ production from FNAL E866. Right panel: Comparison of
E866 $\bar d - \bar u$ data with calculations from various 
models~\cite{towell01}.}
\label{fig:1}       % Give a unique label
\end{figure}
Many theoretical models, including meson cloud model, chiral-quark
model, Pauli-blocking model, instanton model, chiral-quark soliton model,
and statistical model, have been proposed to explain the $\bar d/ \bar u$
asymmetry. For recent reviews, see~\cite{kumano98,garvey02}. 
These models can describe the $\bar d - \bar u$ data very well, 
as shown in Fig. 1. However, they all have difficulties explaining 
the $\bar d / \bar u$ data at large $x$ ($x>0.2$).
The new 120 GeV Fermilab Main Injector 
and the proposed 50 GeV Japanese Hadron Facility present opportunities for
extending the $\bar d/ \bar u$ measurement to larger $x$ ($0.25 < x < 0.7$).

Models in which virtual mesons are admitted as degrees of freedom
have implications that extend beyond the $\bar d, \bar u$
flavor asymmetry addressed above.
They create hidden strangeness in the nucleon via such virtual processes as
$p \to \Lambda + K^+, \Sigma + K$, etc.
Such processes are of considerable interest as they imply different $s$ and
$\bar s$ parton distributions in the nucleon, a feature not found in gluonic
production of $s \bar s$ pairs.

A difference between the $s$ and $\bar s$ distribution can be made manifest
by direct measurements of the $s$ and $\bar s$ parton distribution functions
in neutrino DIS. A fit to the CDHS neutrino charged-current
inclusive data together with  charged lepton DIS data found evidence for
$\int_0^1 s(x) dx > \int_0^1 \bar s(x) dx$~\cite{barone00}. However, 
an analysis~\cite{zeller02a} of 
the recent CCFR and NuTeV $\nu (\bar \nu) N \to \mu^+ \mu^- x$ dimuon 
production data~\cite{goncharov01} 
favored $\int_0^1 s(x) dx < \int_0^1 \bar s(x) dx$ 
($\int_0^1 (s(x) - \bar s(x)) dx = -0.0027 \pm 0.0013$). To better
determine the $s/\bar s$ asymmetry, an NLO analysis is currently 
underway~\cite{olness03}.
Violation of the $s/\bar s$ symmetry would have impact on the recent 
extraction~\cite{zeller02b}
of sin$^2\theta_W$ from the CCFR/NuTeV $\nu N$ scattering data.

Asymmetry in the $s, \bar s$ distributions can also be revealed
in the measurements of the
strange quark's contribution 
to the nucleon's electromagnetic and axial form factors.
These ``strange'' form
factors can be measured in neutrino elastic scattering~\cite{garvey93}
from the nucleon, or by selecting the parity-violating component of
electron-nucleon elastic scattering. Two completed parity-violating
experiments~\cite{spayde00,aniol01} 
suggest small contributions of strange quarks to nucleon
form factors. Several new experiments are underway at JLab and MAMI
to measure parity-violating asymmetry at various kinematic regions.
\subsection{Flavor structure of polarized nucleon sea}
The flavor structure and the
spin structure of the nucleon sea are closely connected. Many
theoretical models originally proposed to explain
the $\bar d / \bar u$ flavor asymmetry also have specific implications for
the spin structure of the nucleon sea. In the meson-cloud model, for example,
a quark would undergo a spin flip upon an
emission of a pseudoscalar meson ($
u^\uparrow \to \pi^\circ (u \bar u, d \bar d) + u^\downarrow,~u^\uparrow
\to \pi^+ (u \bar d) + d^\downarrow,~u^\uparrow \to K^+ + s^\downarrow$,
etc.). The antiquarks ($\bar u, \bar d, \bar s$) are
unpolarized ($\Delta \bar u = \Delta \bar d = \Delta \bar s = 0$)
since they reside in spin-0
mesons. The strange quarks ($s$), on the other hand, would have a negative
polarization.

In the chiral-quark soliton model~\cite{diakonov96,wakamatsu98},
the polarized isovector distributions
$\Delta \bar u(x) - \Delta \bar d(x)$ appears in leading-order ($N_c^2$)
in a $1/N_c$ expansion, while the unpolarized isovector distributions
$\bar u(x) - \bar d(x)$ appear in next-to-leading order ($N_c$).
Therefore, this model predicts a large flavor asymmetry for the polarized sea
$[\Delta \bar u (x) - \Delta \bar d(x)] > [\bar d(x) - \bar u(x)]$.

The HERMES collaboration
has recently reported 
the extraction of $\Delta \bar u(x)$, $\Delta \bar d(x)$, and
$\Delta \bar s(x) (=\Delta s(x))$ using polarized semi-inclusive
DIS (SIDIS) data~\cite{hermes03a}. 
Although the statistics are still limited, the
HERMES results for $\Delta \bar u, \Delta \bar d, \Delta
\bar u - \Delta \bar d$, as shown in Fig. 2, are all 
consistent with being zero. In particular, there is no evidence for a large
positive $\Delta \bar u(x) - \Delta \bar d(x)$ asymmetry as was
predicted~\cite{dressler00} by the chiral quark soliton model.
Figure 2 also shows that $\Delta s$ tends 
to be positive, in contrast to the predictions of a negative polarization 
of the strange sea in the analysis of inclusive DIS and hyperon decay 
data assuming SU(3) symmetry. However, the HERMES result of $\Delta s =
0.03 \pm 0.03 \pm 0.01$ over $0.023 < x < 0.3$ is not in disagreement with 
the inclusive DIS result of 
$(\Delta s + \Delta \bar s)/2 \simeq -0.02$~\cite{adeva98}.

\begin{figure}
\resizebox{0.45\textwidth}{!}{%
  \includegraphics{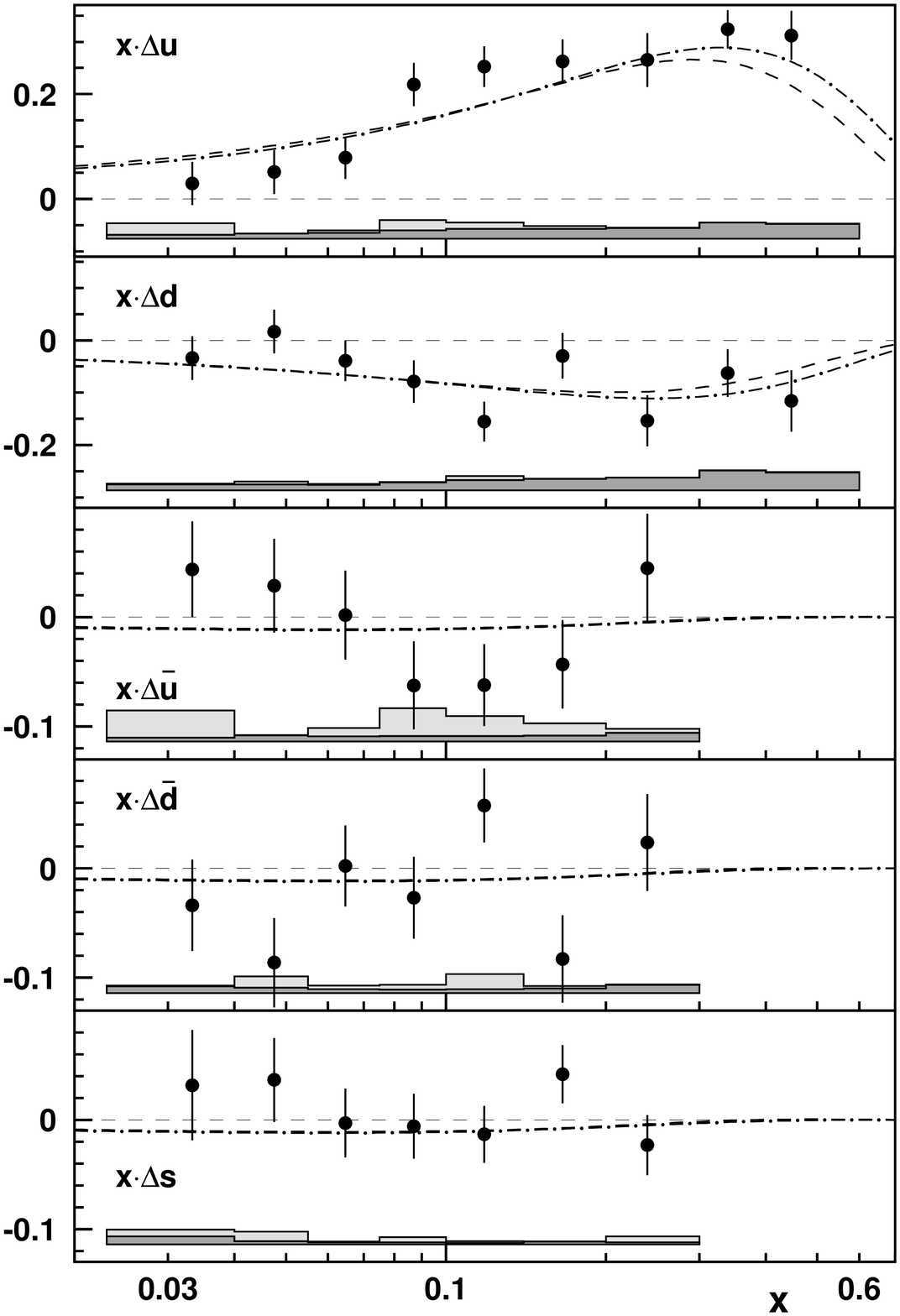}
}
\caption{Quark and antiquark polarizations extracted from the HERMES
SIDIS data~\cite{hermes03a}.}
\label{fig:2}       % Give a unique label
\end{figure}
Another promising technique for measuring sea-quark polarization is
$W$-boson production~\cite{bunce00} at RHIC.
The longitudinal single-spin asymmetry for $W$ production in polarized
$ p + p \to W^{\pm} + x$ gives a direct measure of
sea-quark polarization. The RHIC $W$-production and
the HERMES SIDIS measurements are clearly complementary tools for
determining polarized sea quark distributions.

\subsection{$d/u$ ratio at large $x$}

Another quantity related to the flavor symmetry of the
proton is the $d/u$ ratio at large $x$. Assuming SU(2)$_{spin} \times$
SU(2)$_{flavor}$ symmetry, the proton wave function is given as
\begin{eqnarray}
|p>\uparrow & = & \frac{1}{\sqrt{2}} u\uparrow (ud)_{S=0,S_Z=0} 
+ \frac{1}{\sqrt{18}}u\uparrow (ud)_{S=1,S_Z=0}
- \frac{1}{3} u\downarrow (ud)_{S=1,S_Z=1} \nonumber \\
&  & -\frac{1}{3} d\uparrow
(uu)_{S=1,S_Z=0} + \frac{\sqrt{2}}{3} d\downarrow (uu)_{S=1,S_Z=1}
\label{Eq:proton}
\end{eqnarray}
\noindent The neutron wave function is readily obtained from 
$u \leftrightarrow d$ interchange. In nature, the 
SU(2)$_{spin} \times$ SU(2)$_{flavor}$ symmetry is clearly broken, as
evidenced by the large $N-\Delta$ mass splitting. The dynamic origins
of this symmetry breaking remains unclear. Close and 
Carlitz~\cite{close73,carlitz75} argued that
the dominance of the $S=0$ diquark configuration over the $S=1$ configuration
would account for the $N-\Delta$ mass splitting as well as the SU(2) $\times$
SU(2) symmetry breaking. An alternative suggestion, based on perturbative 
QCD, was offered by Farrar and Jackson~\cite{farrar75}. 
They pointed out that the 
spin-aligned diquark configuration with $S_Z=1$ is suppressed since only 
longitudinal gluons can be exchanged. A similar result was also obtained
by Brodsky et al.~\cite{brodsky95} 
using counting rule argument. It is straightforward to show
that in the $x \to 1$ limit, the different models predict the folllowing
values for various ratios:

\begin{itemize}

\item SU(2)$_{spin} \times$ SU(2)$_{flavor}$ symmetry:~~~
$\frac{d}{u} = \frac{1}{2},~\frac{\Delta u}{u} = \frac{2}{3},~ 
\frac{\Delta d}{d} = - \frac{1}{3},~\frac{F^n_2}{F^p_2} = \frac{2}{3}$.

\item $S=0$ diquark dominance:~~~~~~~~~~~~~~~~~
$\frac{d}{u} = 0,~\frac{\Delta u}{u} = 1,~ 
\frac{\Delta d}{d} = - \frac{1}{3},~\frac{F^n_2}{F^p_2} = \frac{1}{4}$.

\item $S_Z = 0$ diquark dominance:~~~~~~~~~~~~~~~
$\frac{d}{u} = \frac{1}{5},~\frac{\Delta u}{u} = 1,~ 
\frac{\Delta d}{d} = 1,~~~~\frac{F^n_2}{F^p_2} = \frac{3}{7}$.

\end{itemize}

\noindent The distinct predictions for $F^n_2/F^p_2$ from various
models could be tested against DIS experiments. However, there exist
considerable uncertainties in the extraction of $F^n_2$ from the 
measurement of $F^d_2$. Depending on the treatment of
the nuclear effects in the deuteron, very different values for $F^n_2/F^p_2$
(and $d/u$) were obtained at large $x$~\cite{wally96}. 
It is clearly desirable to
measure $d/u$ without the need to model nuclear effects in the
deuteron. One method is to measure the charge asymmetry of $W$ production 
in $p-\bar p$ collision. Indeed, the CDF data~\cite{cdf98} 
on the $W$ charge asymmetry
have already provided useful constraints on the $d/u$ ratio.

\begin{figure}
\resizebox{1.0\textwidth}{!}{%
  \includegraphics{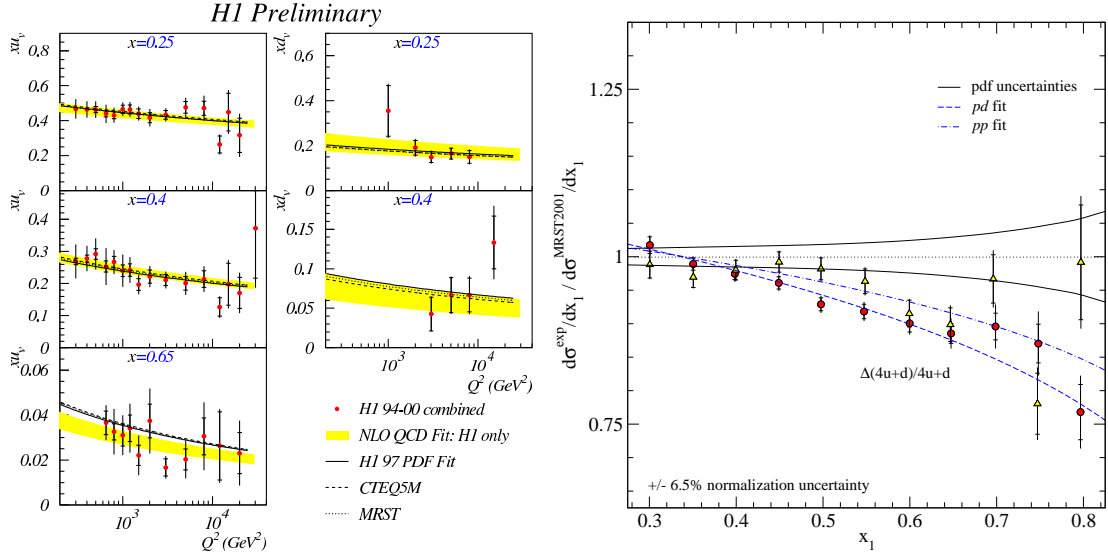}
}
\caption{Left panel: $u$ and $d$ valence quark densities obtained from H1
charged-current measurements~\cite{zhang01}. Right panel: Comparison
of the E866 $pp$ and $pd$ DY cross sections with PDF 
calculations~\cite{webb03}.}
\label{fig:3}       % Give a unique label
\end{figure}
The $d/u$ ratio can also be probed by measuring the $e^- p \to \nu_e x$ and
$e^+ p \to \bar \nu_e x$ charged-current DIS, where the underlying
processes are $e^- u \to \nu_e d$ and $e^+ d \to \bar \nu_e u$,
respectively. The recent H1 charged-current data~\cite{zhang01}, shown 
in Fig. 3, indicate that the $u$ quark density at large $x (x=0.65)$
is smaller than expected from the current PDF parametrization. 
Very recently, the Fermilab
E866/NuSea collaboration reported the absolute Drell-Yan cross
sections of 800 GeV $p + p$ and $p + d$~\cite{webb03}. 
As shown in Fig. 3, the data fall below 
the PDF predictions at large $x$ (up to $x=0.8$). The H1 and the E866
results suggest that $u$ quark density at large $x$ might be smaller
than expected from current PDFs. This clearly would impact on the
$d/u$ ratio at large $x$, as shown in a recent global 
PDF analysis~\cite{tung03}.

The uncertainties involved in the extraction of $F^n_2$ from $F^d_2$ data can
be greatly reduced using the technique of neutron-tagging. A new 
experiment~\cite{bonus03}
has been proposed at the JLab Hall-B to detect $e^- d \to e^- p x$,
where a low-energy recoiled proton will be measured in coincidence with the
$(e,e^{\prime})$ scattering. Using this method, the $F^n_2/F^p_2$ ratio
over the range $0.2 < x < 0.7$ could be determined with small systematic
uncertainties.

\section{Transition from high-$Q^2$ to low-$Q^2$}
\subsection{Quark-hadron duality}
The recent studies at JLab of the spin-averaged and spin-dependent
structure functions at low $Q^2$ region have shed new light on the
subject of quark-hadron duality. Thirty years ago, Bloom and
Gilman~\cite{bloom70}
noticed that the structure functions obtained from deep-inelastic
scattering experiments, where the substructures of the nucleon are
probed, are very similar to the averaged structure functions measured
at lower energy, where effects of nucleon resonances dominate.
This surprising similarity between the resonance electroproduction
and the deep inelastic scattering suggests a common origin for these
two phenomena, called local duality.

Recently, high precision data~\cite{niculescu00}
from JLab have verified the quark-hadron
duality for spin-averaged scattering on proton and deuteron targets.
For $Q^2$ as low as 0.5 GeV$^2$, the resonance data are within 10\%
of the DIS results. When the mean $F_2$ curve from the resonance data is
plotted as a function of the Nachtmann variable,
$\xi = 2x/(1+\sqrt{1+4M^2x^2/Q^2})$, it resembles the $xF_3$ structure
function obtained in neutrino scattering experiments. Since $xF_3$
is a measure of the valence quark distributions, this suggests that the
$F_2$ structure function at low $Q^2$ originates
from valence quarks only.

The study of quark-hadron duality was recently extended
to other structure functions.
Results from HERMES~\cite{hermes03b} show that duality is also observed for
the spin-dependent quantity $A^p_1$. Another recent
result from JLab shows that the nuclear modifications to the unpolarized
structure functions in the resonance region are in surprisingly
good agreement with those measured in DIS~\cite{arrington03}.
\subsection{$\Gamma_1(Q^2)$ at low $Q^2$ and the generalized GDH integral}
The extensive data on $g_1(x,Q^2)$ allow accurate determinations
of the integrals
$\Gamma_1^{p,n}(Q^2) = \int_{0}^{1} g_1^{p,n}(x,Q^2)dx$ for the
proton and the neutron,
as well as $\Gamma_1^p(Q^2) - \Gamma_1^n(Q^2)$.
While the values of $\Gamma_1^p$ and $\Gamma_1^n$ are different from
the predictions of Ellis and Jaffe who assumed SU(3)
flavor symmetry and an unpolarized strange sea, the data are
in good agreement with the prediction of
the Bjorken sum rule.

\begin{figure}
\resizebox{1.0\textwidth}{!}{%
  \includegraphics{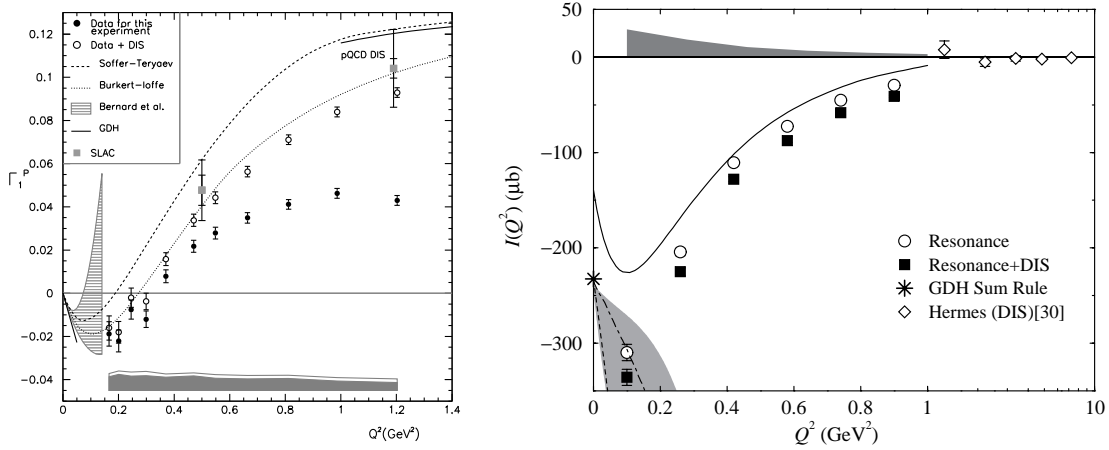}
}
\caption{Left panel: $\Gamma^p_1(Q^2)$ from CLAS~\cite{fatemi03}.
Right panel: Generalized GDH integral from JLab Hall-A 
experiment~\cite{amarian02}.}
\label{fig:4}       % Give a unique label
\end{figure}
How does $\Gamma_1(Q^2)$ evolve as $Q^2 \to 0$? This question
is closely related to the Gerasimov-Drell-Hearn (GDH) sum
rule:

\begin{equation}
\int_{\nu_0}^{\infty} [\sigma_{1/2}(\nu) - \sigma_{3/2}(\nu)] \frac{d\nu}{\nu}
= -\frac{2\pi^2\alpha}{M^2} \kappa^2.
\label{Eq:GDH}
\end{equation}

\noindent The GDH sum rule, based
on general physics principles (causality, unitarity,
Lorentz and gauge invariances) and dispersion relation, relates the total
absorption cross sections of circularly polarized photons on longitudinally
polarized nucleons to the static properties of the nucleons.
In Eq.~\ref{Eq:GDH}, $\sigma_{1/2}$ and $\sigma_{3/2}$ are the photo-nucleon
absorption cross sections of total helicity of $1/2$ and $3/2$, $\nu$ is
the photon energy and $\nu_0$ is the pion production threshold, $M$ is
the nucleon mass and $\kappa$ is the nucleon anomalous magnetic moment.
The GDH integral in Eq.~\ref{Eq:GDH} can be generalized from real photon
absorption to virtual photon absorption with non-zero $Q^2$:

\begin{eqnarray}
I_{GDH}(Q^2) \equiv \int_{\nu_0}^{\infty} [\sigma_{1/2}(\nu,Q^2)
- \sigma_{3/2}(\nu,Q^2)] \frac{d\nu}{\nu} = \frac{16\pi^2\alpha}{Q^2} 
\Gamma_1(Q^2).
\label{Eq:GGDH}
\end{eqnarray}

\noindent Eq.~\ref{Eq:GGDH} shows that
the $Q^2$-dependence of the generalized GDH integral
is directly related to the $Q^2$-dependence of $\Gamma_1$.
The GDH sum rule (Eq.~\ref{Eq:GDH})
predicts $\Gamma_1^p = 0$
at $Q^2=0$ with a negative slope for $d\Gamma_1^p(Q^2)/dQ^2$ and
$\Gamma_1^p$ is known to be
positive at high $Q^2$, therefore,
$\Gamma_1^p(Q^2)$ must become negative at
low $Q^2$.

The GDH integrals at low $Q^2$ have recently been measured
in several experiments at JLab~\cite{amarian02,fatemi03} 
and HERMES~\cite{hermes03c}. Results from a JLab Hall-B 
measurement~\cite{fatemi03} of
$\Gamma_1^p(Q^2)$ are shown in
Fig. 4. These data indeed show that $\Gamma_1^p$ changes
sign around $Q^2 = 0.3$ GeV$^2$. The origin of the sign-change
can be attributed to the competition between $\Delta(1232)$ and
higher nucleon resonances. At the lowest $Q^2$, the $\Delta(1232)$ has a
dominant negative contribution to $\Gamma_1^p$. However, at larger
$Q^2$, higher mass nucleon resonances take over to have a net positive
$\Gamma_1^p$.

Results~\cite{amarian02} from a JLab Hall-A measurement of the generalized
GDH integral for neutron using a polarized $^3$He target are shown in Fig. 4.
In contrast to the proton case, the strong negative
contribution to the GDH integral from the
$\Delta(1232)$ resonance now dominates the entire measured $Q^2$ range.
Future experiments at JLab will extend
the measurements down to $Q^2=0.02$ GeV$^2$ in order to map out
the low $Q^2$ behavior of the neutron and proton generalized GDH integrals.

\section{Novel distribution and fragmentation functions}

In addition to the unpolarized and polarized quark distributions, $q(x,Q^2)$
and $\Delta q(x,Q^2)$, a third quark distribution, called transversity, is the
remaining twist-2 distribution yet to be measured.
This helicity-flip quark distribution,
$\delta q(x,Q^2)$, can be described in quark-parton model 
as the net transverse
polarization of quarks in a transversely polarized nucleon. 
Due to the chiral-odd nature of the transversity distribution, it can not
be measured in inclusive DIS experiments. In order 
to measure $\delta q(x,Q^2)$,
an additional chiral-odd object is required. For example, the double
spin asymmetry, $A_{TT}$, for Drell-Yan cross section
in transversely polarized $p p$ collision, is sensitive to transversity
since $A_{TT} \sim \sum_{i} e_i^2 \delta q_i(x_1) \delta \bar q_i(x_2)$.
Such a measurement could be carried out at RHIC~\cite{bunce00},
although the anticipated effect is small, on the order of $1-2$\%.

Several other methods for measuring transversity have been proposed for
semi-inclusive DIS. In particular, Collins suggested~\cite{collins93}
that a chiral-odd
fragmentation function in conjunction with the chiral-odd transversity
distribution would lead to a single-spin azimuthal asymmetry
in semi-inclusive pion production. 

The HERMES collaboration recently reported~\cite{hermes03d} observation of
single-spin azimuthal asymmetry
for charged and neutral hadron electroproduction. Using
unpolarized positron beam on a longitudinally polarized hydrogen and deuterium
targets, the cross section was found to have a sin$\phi$ dependence
correlating with the target spin direction.
$\phi$ is the azimuthal angle between the pion and the $(e, e^\prime)$
scattering plane.
This Single-Spin-Asymmetries (SSA) can be expressed as the analyzing power
in the sin$\phi$ moment, and the result is shown in Fig. 5.
The sin$\phi$ moment for an unpolarized (U) positron scattered off
a longitudinally (L) polarized target contains two main contributions
\begin{eqnarray}
\langle sin \phi \rangle & \alpha & S_L \frac{2 (2-y)}{Q\sqrt{1-y}}
\sum_{q} e_q^2 x h_L^q(x) H_1^{\bot,q}(z)
+ S_T (1-y) \sum_{q} e_q^2 x h_1^q(x) H_1^{\bot,q}(z),
\label{Eq:ssa1}
\end{eqnarray}
\noindent where $S_L$ and $S_T$ are the longitudinal and transverse components
of the target spin orientation with respect to the virtual photon direction.
For the HERMES experiment with a longitudinally polarized target, the
transverse component is nonzero with a mean value of $S_T \approx 0.15$.
The observed azimuthal asymmetry could be a combined effect of the
$h_1$ transversity and the twist-3 $h_L$ distribution. 
Recently, another mechanism involving a chiral-even 
T-odd Sivers distribution function~\cite{sivers90}
was shown to contribute to azimuthal asymmetry~\cite{brodsky02,collins02}. 
For a longitudinally
polarized target the Collins and the Sivers mechanisms can not be 
distinguished.

\begin{figure}
\resizebox{0.45\textwidth}{!}{%
  \includegraphics{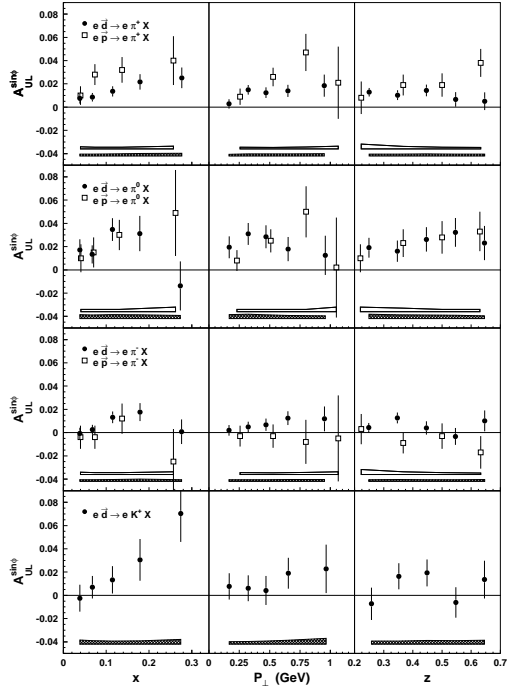}
}
\caption{Analyzing power in the sin$\phi$ moment from HERMES~\cite{hermes03c}.}
\label{fig:5}       % Give a unique label
\end{figure}
If the azimuthal asymmetry observed by HERMES is indeed caused by the
$h_1$ transversity, a much larger asymmetry is expected for a transversely
polarized target. 
The HERMES and COMPASS collaborations have 
collected polarized SIDIS using transversely polarized hydrogen and
$^6$LiD targets, respectively. These data would shed much light on the
origins of the SSA and could also disentangle the Sivers effect from the 
Collins effect. The Collins effect has a sin$(\phi^l_h + \phi^l_s)$ dependence
while the Sivers effect is proportional to sin$(\phi^l_h - \phi^l_s)$,
where $\phi^l_s = \phi_s - \phi^l$ is the angle between target spin and
the lepton scattering plane. For longitudinally polarized target 
$\phi^l_s = 0$ and the two effects have identical $\phi$ dependence. For 
transversely polarized target, however, $\phi^l_s \neq 0$ and the 
two effects can be separated.

The Collins fragmentation function represents a correlation between the 
quark's transverse spin and the transverse momentum of the leading hadron
formed in the fragmentation process. The Sivers distribution function 
reflects the correlation between the quark's transverse spin and its
transverse momentum within the proton. It 
has been shown~\cite{anselmino95,anselmino98} that both the
Collins and the Sivers effects can contribuite to the analysing power
$A_N$ observed in the Fermilab E704 
$p\uparrow p \to \pi x$ reaction~\cite{adams91}.
Very recently, $A_N$ was measured~\cite{bland02}
at RHIC at a much higher energy of 
$\sqrt s =$ 200 GeV using transversely polarized proton beams. The RHIC
data could provide new information on the Collins and Sivers functions.

\section{Generalized parton distributions}

There has been intense theoretical and experimental activities in recent years
on the subject of Generalized Parton Distribution (GPD). In the Bjorken scaling
regime, exclusive leptoproduction reactions can be factorized into a 
hard-scattering
part calculable in QCD, and a non-perturbative part parameterized by the GPDs.
The GPD takes into account dynamical correlations between partons with 
different momenta. In addition to the dependence on $Q^2$ 
and $x$, the GPD also depends on
two more parameters, the skewedness $\xi$ and the momentum transfer to the
baryon, $t$. Of particular interest is the connection 
between GPD and the nucleon's
orbital angular momentum~\cite{ji97}.

The deeply virtual Compton scattering (DVCS), in which an energetic photon is
produced in the reaction $e p \to e p \gamma$, is most 
suitable for studying GPD.
Unlike the exclusive meson productions, DVCS avoids the complication 
associated with
mesons in the final state and can be cleanly interpreted in terms of GPDs.
An important experimental challenge, however, is to separate the relatively
rare DVCS events from the abundant electromagnetic Bethe-Heitler (BH) 
background. From the collision
of 800 GeV protons with 27.5 GeV positrons, both the ZEUS~\cite{chekanov03}
and the H1~\cite{adloff01} collaborations at DESY observed an excess of
$e^+ + p \to e^+ + \gamma + p$ events in a kinematic region where
the BH cross section is largely suppressed. The excess events were
attributed to the DVCS process and the ZEUS collaboration
further determined~\cite{chekanov03} the DVCS cross section over the
kinematic range $5 < Q^2 < 100$ GeV$^2$, $40 < W < 140$ GeV. Both the 
$W$ and $Q^2$ depndences of the ZEUS DVCS cross section data are well
described by calculations based on GPD and on the color-dipole model.

At lower c.m. energies, the HERMES~\cite{airapetian01} and the
CLAS~\cite{stepanyan01} collaborations
observed the interference between the DVCS and the BH processes,
which manifests itself as a pronounced sin$\phi$ 
azimuthal asymmetry correlated with
the beam helicity. 
Another observable
sensitive to the interference between the DVCS and the BH processes
is the azimuthal asymmetry between unpolarized $e^+$ and 
$e^-$ beams. In contrast
to the Beam Spin Asymmetry (BSA) which is sensitive to the imaginary part of
the DVCS amplitudes, the Beam Charge Asymmetry (BCA) is probing the real part
of the DVCS amplitudes~\cite{diehl97}. Analysis of the HERMES $e^-$ data
in 98-99 and the
$e^+$ data in 99-00 has shown a positive effect for BSA~\cite{bianchi03}.

QCD factorization was proved to be valid
for exclusive meson production with longitudinal 
virtual photons~\cite{collins97}.
Such factorization allowed new means to extract the unpolarized and polarized
GPD. In particular, unpolarized GPDs can be measured
with exclusive vector meson production, while polarized GPDs can be probed via
exclusive pseudoscalar meson production. A broad program of DVCS and
hard exclusive processes has been proposed~\cite{burkert03} for 
the 12 GeV upgrade at JLab.

%%%%%%%%%%%%%%%%%%%%%%%%%%%%%%%%%%%%%%%%%%%%%%%%
%% BACKMATTER
%%%%%%%%%%%%%%%%%%%%%%%%%%%%%%%%%%%%%%%%%%%%%%%%

\begin{theacknowledgments}
I would like to thank V. Burkert, J. P. Chen, C. Keppel, N. Makins,
and W. K. Tung for
helpful discussion.
\end{theacknowledgments}

%%%%%%%%%%%%%%%%%%%%%%%%%%%%%%%%%%%%%%%%%%%%%%%%
%% You may have to change the BibTeX style below, depending on your
%% setup or preferences.
%%
%% If the bibliography is produced without BibTeX comment out the
%% following lines and see the aipguide.pdf for further information.
%%
%% For The AIP proceedings layouts use either
%%%%%%%%%%%%%%%%%%%%%%%%%%%%%%%%%%%%%%%%%%%%

\bibliographystyle{aipproc}   % if natbib is available
%\bibliographystyle{aipprocl} % if natbib is missing

%%%%%%%%%%%%%%%%%%%%%%%%%%%%%%%%%%%%%%%%%%%
%% You probably want to use your own bibtex database here
%%%%%%%%%%%%%%%%%%%%%%%%%%%%%%%%%%%%%%%%%%%
\bibliography{sample}

%%%%%%%%%%%%%%%%%%%%%%%%%%%%%%%%%%%%%%%%%%%
%% Just a reminder that you may have to run bibtex
%% All of it up to \end{document} can be removed
%% if you don't like the warning.
%%%%%%%%%%%%%%%%%%%%%%%%%%%%%%%%%%%%%%%%%%%
\IfFileExists{\jobname.bbl}{}
 {\typeout{}
  \typeout{******************************************}
  \typeout{** Please run "bibtex \jobname" to optain}
  \typeout{** the bibliography and then re-run LaTeX}
  \typeout{** twice to fix the references!}
  \typeout{******************************************}
  \typeout{}
 }

\end{document}